\documentclass[letterpaper,twocolumn,prx,aps,showpacs,groupaddress,floatfix]{revtex4-1}
\usepackage[latin1]{inputenc}
\usepackage{bm}
\usepackage[usenames]{color}
\usepackage{multirow}
\usepackage{amssymb}
\usepackage{amsbsy}
\usepackage{amsmath}
\usepackage{stmaryrd}
\usepackage{graphicx}
\usepackage{epsfig}
\usepackage{placeins}
\usepackage{ulem}
\usepackage{bbold}
\usepackage{braket}
\usepackage{blindtext}
\usepackage[colorlinks,linkcolor=blue,citecolor=blue,urlcolor=blue]{hyperref}

\makeatletter

\begin{document}

\title{Combining dynamical quantum typicality and numerical linked cluster expansions}

\author{Jonas Richter}
\email{jonasrichter@uos.de}
\affiliation{Department of Physics, University of Osnabr\"uck, D-49069 Osnabr\"uck, Germany}

\author{Robin Steinigeweg}
\email{rsteinig@uos.de}
\affiliation{Department of Physics, University of Osnabr\"uck, D-49069 Osnabr\"uck, Germany}

\date{\today}


\begin{abstract}
We demonstrate that numerical linked cluster expansions (NLCE) yield a 
powerful approach to calculate time-dependent correlation 
functions for quantum many-body systems in one dimension. As a paradigmatic 
example, we study the dynamics of the spin current in the spin-$1/2$ XXZ  chain 
for different values of anisotropy, as well as in the presence of an 
integrability-breaking next-nearest neighbor interaction. For short to 
intermediate time scales, we unveil that NLCE yields a convergence towards 
the thermodynamic limit already for small cluster sizes, which is much faster 
than in direct calculations of the autocorrelation function for systems with 
open or periodic boundary conditions. Most importantly, we show that the range 
of accessible cluster sizes in NLCE can be extended by evaluating the 
contributions of larger clusters by means of a pure-state approach based on the 
concept of dynamical quantum typicality (DQT). Even for moderate computational 
effort, this combination of DQT and NLCE provides a competitive alternative to 
existing state-of-the-art techniques, which may be applied in higher dimensions 
as well.
\end{abstract}

\maketitle


\section{Introduction}
Unraveling the complex dynamics of interacting quantum many-body systems is a 
central area of research of modern experimental and theoretical physics  
\cite{eisert2015}. On the one hand, fascinating experiments with cold atoms 
\cite{Bloch2012, Langen2015} and trapped ions \cite{Blatt2012} nowadays open 
the possibility to explore the unitary time evolution of closed quantum systems 
for a variety of tailored Hamiltonians and initial states. On the other hand, 
theoretical studies of interacting many-body systems are challenging as well, 
and analytic solutions are comparatively rare \cite{essler2016, Alvaredo2016, 
Bertini2016}. Nevertheless, much progress has been made due to the development 
of sophisticated numerical tools including, e.g., dynamical mean field theory 
\cite{Aoki2014}, Krylov subspace approaches \cite{Nauts1983, Long2003}, quantum 
Monte-Carlo \cite{Goth2012}, classical representations in phase space  
\cite{Wurtz2018} or networks \cite{Schmitt2018}, as well as innovative machine 
learning implementations \cite{Carleo2017}, to name just a few. While each of 
these methods certainly has its own specific strengths and drawbacks, their 
combination provides a comprehensive picture for a wide range of physical 
situations. Particularly for one-dimensional systems, the time-dependent 
density matrix renormalization group (tDMRG), including related methods based 
on matrix product states, are powerful techniques to study the dynamical 
properties of quantum systems practically in the thermodynamic limit 
\cite{schollwoeck20052011, Vidal2004, White2004}. However, since such 
approaches rely on an efficient compression of the wave function, they are 
generally limited by the build-up of entanglement which in turn restricts the 
maximum time reachable in simulations \cite{Sirker2009, Kennes2016}.  

Among the numerous methods available, exact diagonalization (ED) is arguably  
the most versatile approach. It can be applied to any finite-dimensional 
Hamiltonian, observable, and initial state. Moreover, it allows the calculation 
of quantum dynamics for arbitrarily long time scales at all temperatures 
\cite{Sandvik2010, Narozhny1998, Heidrichmeisner2003, Steinigeweg2011, 
Herrera2015, Schmitt2018_2}. However, ED is generally limited to rather small 
Hilbert-space dimensions, and this dimension grows exponentially fast for 
many-body problems. Substantially larger Hilbert spaces can be treated by, 
e.g., the concept of dynamical quantum typicality (DQT), where static and 
dynamic expectation values are evaluated on the basis of pure quantum states 
which mimic the statistical ensemble \cite{Gemmer2004, Popescu2006, 
Goldstein2006, Reimann2007, bartsch2009, Hams2000, iitaka2003, sugiura2013, 
elsayed2013, monnai2014, steinigeweg2014}. Both, ED and DQT, will 
be important building blocks of the present work. To be concrete, we will 
combine ED and DQT within the framework of so-called numerical linked cluster 
expansions (NLCE) \cite{Tang2013}. While NLCE have been originally introduced 
to study quantum systems in equilibrium \cite{Rigol2006, Rigol2007} (see also 
Refs.\ \cite{Khatami2011, Ixert2015}), they also proved to be a useful approach 
to calculate entanglement entropies \cite{Kallin2013} and to predict 
steady-state properties in quantum quench scenarios \cite{Rigol2014, 
Wouters2014, Mallayya2017} and driven-dissipative systems \cite{Biella2018}. 
More recently, NLCE have been successfully employed to access the full time 
evolution of observables resulting from a quench, with the initial state being 
either a (pure) product state \cite{White2017, Sanchez2018} or also a (mixed) 
thermal density matrix \cite{Mallayya2018}. 

In this context, the present paper demonstrates that NLCE also yield a  
powerful approach to calculate time-dependent current autocorrelations for 
one-dimensional quantum spin models. We particularly unveil that, on short to 
intermediate time scales, NLCE can outperform standard finite-size scaling in 
systems with open or periodic boundary conditions. Moreover, we show that NLCE 
can be significantly improved if the contributions of larger clusters are 
evaluated by means of DQT. This combination of NLCE and DQT provides a 
competitive alternative to existing state-of-the-art techniques operating in 
the thermodynamic limit.

This paper is structured as follows. In Sec. \ref{model} we first introduce the 
model. In Sec. \ref{numerical_approach} we then give an overview over selected 
numerical methods, including ED, DQT, and NLCE, and particularly discuss the 
combination of DQT and NLCE. In Sec.\ \ref{application} this combination is 
applied and compared to other numerical approaches. Finally, we summarize 
and conclude in Sec.\ \ref{conclusion}.


\section{Model} \label{model}
As a paradigmatic example, we consider the spin-$1/2$ XXZ chain, described by  
the Hamiltonian
\begin{equation}\label{Eq::Ham}
{\cal H} = J\sum_{l=1}^{L-1,L} \left(S_l^x S_{l+1}^x + S_l^y S_{l+1}^y + \Delta 
S_l^z S_{l+1}^z \right)\ , 
\end{equation}
where $S_l^{\mu}$ $(\mu = x,y,z)$ are spin-$1/2$ operators at site $l$, $J > 0$ 
is the antiferromagnetic coupling constant, and $\Delta$ denotes the anisotropy 
in the $z$ direction. Moreover, $L$ is the number of sites, and the sum in Eq.\ 
\eqref{Eq::Ham} runs up to $L - 1$ ($L$) if one is interested in open 
(periodic) boundary conditions ($S_{L+1}^\mu = S_1^\mu$). While this model is 
integrable in terms of the Bethe ansatz \cite{Zotos1999, Klumper2002}, this 
integrability can be broken, e.g., by an additional next-nearest neighbor 
interaction $\Delta' \neq 0$ 
\cite{Heidrichmeisner2003, Steinigeweg2013, Richter2018}, 
\begin{equation} \label{Eq::DeltaP}
{\cal H} \to {\cal H} +  J\sum_{l=1}^{L-2,L} \Delta' S_l^z S_{l+2}^z\ . 
\end{equation}
Since the total magnetization $M = \sum_l S_l^z$ is conserved for all values  
of $\Delta$ and $\Delta'$, the spin current is well-defined and has the 
well-known form
\cite{heidrichmeisner2007}
\begin{equation} \label{Eq::Cur}
j = J\sum_{l=1}^{L-1,L} \left(S_l^x S_{l+1}^y - S_l^y S_{l+1}^x \right)\ ,   
\end{equation}
which also depends on the specific boundary condition chosen. In this paper, we 
explore the time dependence of the current autocorrelation function 
\begin{equation}\label{Eq::JJ}
\langle j(t) j \rangle_\text{eq} = \text{Tr}[j(t) j \rho_\text{eq}]\ ,
\end{equation}
where $\rho_\text{eq} = e^{-\beta {\cal H}}/{\cal Z}_\text{eq}$ is the  
canonical ensemble at inverse temperature $\beta = 1/T$, and $j(t) = e^{i{\cal 
H}t} j e^{-i{\cal H}t}$. Within linear response theory (LRT), $\langle j(t) j 
\rangle_\text{eq}$ is directly related to transport properties via the Kubo 
formula. For earlier studies of current autocorrelations in the spin-$1/2$ XXZ 
chain see, e.g., Refs.\ \cite{Sirker2009, steinigeweg2014, Steinigeweg2009, 
Karrasch2012, Karrasch2013, Karrasch2015, steinigeweg2015}. Note further that 
there exist of course also other approaches to transport in low-dimensional 
quantum spin systems apart from LRT \cite{Michel2008, Prosen2009, 
Ljubotina2017}. 


\section{Numerical approach} \label{numerical_approach}
Before discussing NLCE below in detail, it is instructive to briefly reiterate 
how to calculate time-dependent correlation functions such as $\langle j(t) j 
\rangle_\text{eq}$ by ED and DQT.

\subsection{Exact diagonalization}

Upon diagonalizing ${\cal H}$ for finite $L$, the full knowledge of eigenstates 
and eigenenergies in principle allows for the computation of all static and 
dynamic properties. In this context, $\langle j(t) j \rangle_\text{eq}$ is 
conveniently written in a spectral representation,
\begin{equation} \label{Eq::ED}
\langle j(t) j \rangle_\text{eq} = \frac{1}{{\cal Z}_\text{eq}}\sum_{m,n} 
e^{-\beta E_m} e^{-i(E_n-E_m)t} |\hspace{-0.1cm}\bra{m} j  
\ket{n}\hspace{-0.1cm}|^2 \ , 
\end{equation}
where the sum runs over all eigenstates $\ket{m}$, $\ket{n}$ of ${\cal H}$ with 
respective eigenenergies $E_m$, $E_n$. Due to the exponential growth of the 
Hilbert space, however, ED is limited to rather small system sizes. This 
limitation becomes particularly severe in the case of open boundary conditions 
where translation symmetry cannot be exploited. Nevertheless, ED for small 
systems will be a major cornerstone for NLCE.

\subsection{Dynamical quantum typicality}

For system sizes outside the range of ED, the method of DQT \cite{Gemmer2004, 
Popescu2006, Goldstein2006, Reimann2007, bartsch2009, Hams2000, iitaka2003, 
sugiura2013, elsayed2013, monnai2014, steinigeweg2014} has been established as 
a very useful numerical approach. This method relies on the fact that even a 
single pure state can have the same properties as the full statistical 
ensemble. Specifically, $\langle j(t) j \rangle_\text{eq}$ can be written as a 
simple scalar product with two (auxiliary) pure states  $\ket{\psi_\beta(t)}$ 
and $\ket{\varphi_\beta(t)}$, \cite{elsayed2013, steinigeweg2014, steinigeweg2015}
\begin{equation}\label{Eq::DQT}
\langle j(t) j \rangle_\text{eq} = \frac{\bra{\psi_\beta(t)}j 
\ket{\varphi_\beta(t)}}{\braket{\psi_\beta(0)|\psi_\beta(0)}} + \epsilon\ , 
\end{equation}
$\ket{\psi_\beta(t)}=e^{-i{\cal H}t}e^{-\beta {\cal H}/2}\ket{\psi}$,  
$\ket{\varphi_\beta(t)} =  e^{-i{\cal H}t} j e^{-\beta {\cal 
H}/2}\ket{\psi}$, and the reference pure state $\ket{\psi}$ is drawn at random 
from the full Hilbert space according to the unitary invariant Haar measure 
\cite{bartsch2009}. Importantly, the statistical error $\epsilon = 
\epsilon(\ket{\psi})$ has zero mean, $\bar{\epsilon} = 0$, and its standard 
deviation scales as $\sigma(\epsilon) \propto 1/\sqrt{d_\text{eff}}$, where 
$d_\text{eff} = {\cal Z}_\text{eq}/e^{-\beta E_0}$ is the effective dimension of 
the Hilbert space and $E_0$ is the ground-state energy of ${\cal H}$ 
\cite{bartsch2009, Hams2000, elsayed2013, steinigeweg2014, steinigeweg2015}.  
Thus, $\sigma(\epsilon)$ decreases exponentially with increasing $L$ and, for 
many practical purposes, $\epsilon$ can be neglected for medium-sized systems 
already (especially for $\beta \to 0$). However, if one wants to improve the 
accuracy of the DQT approximation even further, it is of course always possible 
to evaluate (nominator and denominator of) Eq.\ \eqref{Eq::DQT} as an average 
over $N_S$ independent realizations of the random pure state $\ket{\psi}$ 
\cite{iitaka2003, Rousochatzakis2018}. In fact, such an averaging turns 
out to be important when NLCE is combined with DQT as done below. 

The main advantage of Eq.\ \eqref{Eq::DQT} comes from the fact that the time  
evolution of pure states can be generated by iteratively solving the 
Schr\"odinger equation. To this end, various sophisticated methods are 
available such as, e.g., Trotter decompositions \cite{deReadt2006}, 
Chebychev polynomials \cite{dobrovitski2003, weisse2006}, Krylov subspace 
techniques \cite{Nauts1983, Varma2017}, and Runge-Kutta schemes 
\cite{elsayed2013, steinigeweg2014}. A unifying feature of all these methods is 
that they essentially require the calculation of matrix-vector products, which 
can be implemented both time- and memory-efficient due to the sparseness of the 
involved operators. Thus, no diagonalization of ${\cal H}$ is needed and Eq.\ 
\eqref{Eq::DQT} can be evaluated for Hilbert-space dimensions substantially 
larger compared to ED.

\subsection{Numerical linked cluster expansions}

Let us now come to NLCE. Note that we intentionally refrain from a general 
introduction to NLCE, for detailed explanations see, e.g., Ref.\ 
\cite{Tang2013}. Instead, we choose to sketch more specifically how NLCE can be 
used to obtain current autocorrelations in a one-dimensional geometry. Within 
NLCE, an extensive quantity per lattice site is calculated as the sum of 
contributions from all connected clusters which can be embedded on the lattice,
\begin{equation}\label{Eq::NLCE}
\langle j(t) j \rangle_\text{eq}/L = \sum_{c} {\cal L}_c W_c(t)\ , 
\end{equation}
where $W_c(t)$ is the weight of cluster $c$ with multiplicity ${\cal L}_c$. 
While the identification of all linked clusters for a given (arbitrary) lattice 
can be a cumbersome procedure, this identification becomes straightforward in 
one dimension. Given an infinitely long and translational-invariant chain, the 
linked clusters are just chains as well, which comprise a certain (finite) 
number of sites. Moreover, for a fixed cluster size, there exists only a single 
topologically distinct cluster $c$ (since any translation of $c$ is just 
equivalent to $c$), cf.\ \cite{Note}. Therefore, we have ${\cal L}_c = 1$ in 
Eq.\ \eqref{Eq::NLCE} and we can identify the cluster index $c$ as the number 
of sites in the respective cluster. The weights $W_c(t)$ in Eq.\ 
\eqref{Eq::NLCE} are calculated by the so-called inclusion-exclusion principle, 
\begin{equation}\label{Eq::Weight1}
W_c(t) = \langle j(t) j \rangle_\text{eq}^{(c)} - \sum_{s \subset c} W_s(t)\ , 
\end{equation}
where $\langle j(t) j \rangle_\text{eq}^{(c)}$ denotes the current 
autocorrelation evaluated on the cluster $c$, and the sum runs over 
all subclusters of $c$. Due to the definition of $j$ in Eq.\ \eqref{Eq::Cur}, 
the smallest nontrivial cluster which needs to be considered is a single bond 
connecting just two lattice sites. The weight of this cluster then follows as 
$W_2(t) = \langle j(t) j \rangle_\text{eq}^{(2)}$, since there are no 
subclusters in this case. A cluster of length $c = 3$ obviously has only 
two linked subclusters of length $c = 2$, i.e., $W_3(t) = \langle j(t) j 
\rangle_\text{eq}^{(3)} - 2 W_2(t)$. Generalizing this scheme, the weight for 
$c\geq3$ reads
\begin{equation}\label{Eq::Weight}
W_c(t) = \langle j(t) j \rangle_\text{eq}^{(c)} - \sum_{s = 2}^{c-1} (c-s+1) 
W_s(t)\ . 
\end{equation}
To summarize, $\langle j(t)  j \rangle_\text{eq}/L$ in the thermodynamic limit 
$L \to \infty$ is calculated as a sum over weights $W_c(t)$ and the calculation 
of these weights requires the evaluation of $\langle j(t) j 
\rangle_\text{eq}^{(c)}$ on clusters with increasing size $c = 2,3,\dots$ and 
open boundary conditions.

In practice, however, it is only possible to consider contributions of clusters 
which are small enough to be treated numerically, and the sum in Eq.\ 
\eqref{Eq::NLCE} eventually has to be truncated to a maximum cluster size $C$. 
On the one hand, for NLCE implementations of thermodynamic quantities, a larger 
value of $C$ often improves the convergence of the expansion down to lower 
temperatures \cite{Bhattaram2018}. On the other hand, for time-dependent 
quantities, the value of $C$ will directly correspond to the time scale on 
which the NLCE can yield reliable results, cf.\ Ref.\ \cite{White2017}. Thus, 
it is generally highly desirable to include clusters as large as possible. 
In this paper, we demonstrate that NLCE can be significantly improved if it is 
combined with DQT in order to evaluate larger clusters outside the range of ED.

Remarkably, in the present one-dimensional situation, it is straightforward to 
show that a truncation of Eq.\ \eqref{Eq::NLCE} to order $C$ takes on the 
simple form
\begin{equation}\label{Eq::Simple}
\sum_{c=2}^C W_c(t) =  \langle j(t) j \rangle_\text{eq}^{(C)} - \langle j(t) j 
\rangle_\text{eq}^{(C-1)}\ , 
\end{equation}
and is just the difference between the two largest clusters $C$ and $C-1$. 
Since this difference might be sensitive to numerical inaccuracies, it is in 
this context convenient to average the DQT calculations over $N_S$ independent 
random states. We here choose $N_S \times 2^c > 5000 \times 2^{17}$, see also 
the table in Appendix \ref{appendix_rs}.
\begin{figure}[tb]
\centering
\includegraphics[width=0.85\columnwidth]{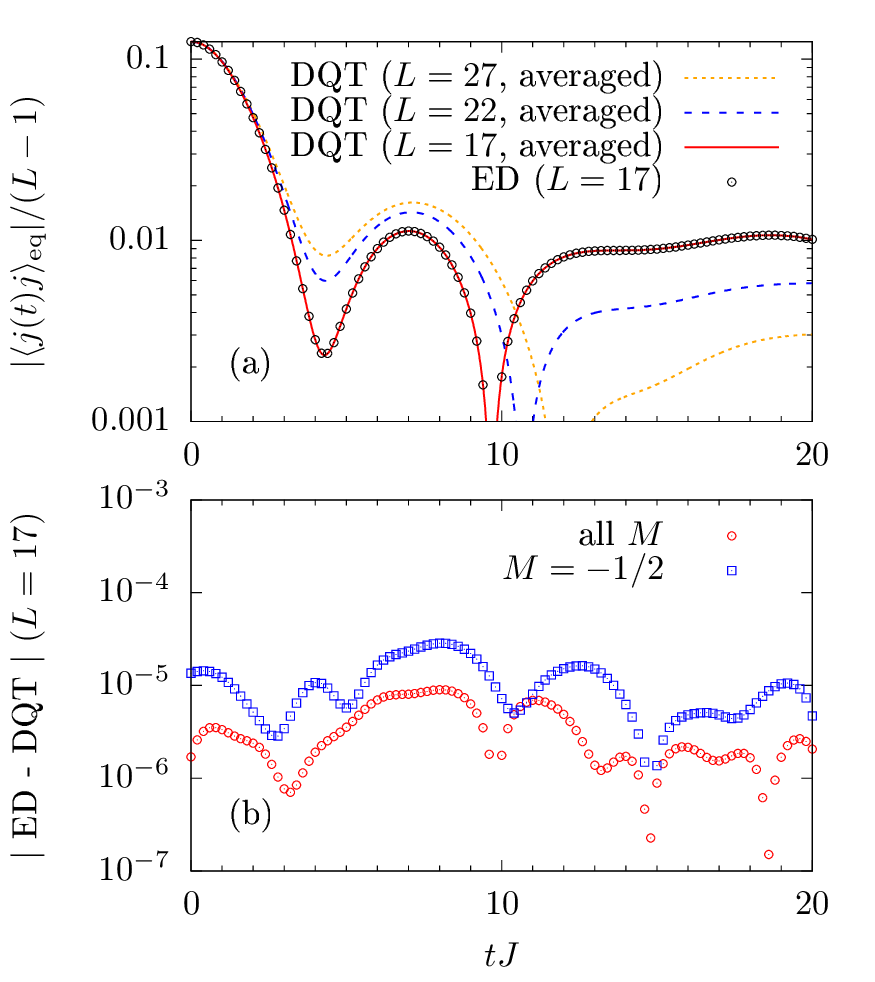}
\caption{(Color online) (a) $|\langle j(t) j \rangle_\text{eq}|/(L-1)$ for $L = 
17$ (ED) and $L = 17,22,27$ (DQT) for systems with open boundary conditions. 
(b) Absolute difference between ED and DQT ($L = 17$). Averaging over all 
magnetization sectors, as done in this paper, improves the accuracy further. 
Other parameters: $\Delta = 1$, $\Delta' = 0$, and $\beta = 0$.}
\label{Fig1}
\end{figure}
%


\section{Application} \label{application}
We start with discussing the accuracy of the pure-state approach 
in Eq.\ \eqref{Eq::DQT}. Thus, Fig.\ \ref{Fig1}~(a) exemplarily shows $|\langle 
j(t) j\rangle_\text{eq}|$ for $\Delta = 1$, $\Delta' = 0$, and $\beta = 0$, 
calculated by ED and DQT for open boundary conditions and $L = 17$. In the 
semilogarithmic plot, one finds that both methods agree nicely with each other, 
indicating that the statistical error $\epsilon$ in Eq.\ \eqref{Eq::DQT} is 
indeed very small. In addition, we depict DQT data for $L = 22, 27$, which 
visualize that finite-size effects become non-negligible already for times $tJ 
\gtrsim 5$. Note that ED is already unfeasible for these $L$ such that no 
comparison is possible.
\begin{figure}[tb]
\centering
\includegraphics[width=0.85\columnwidth]{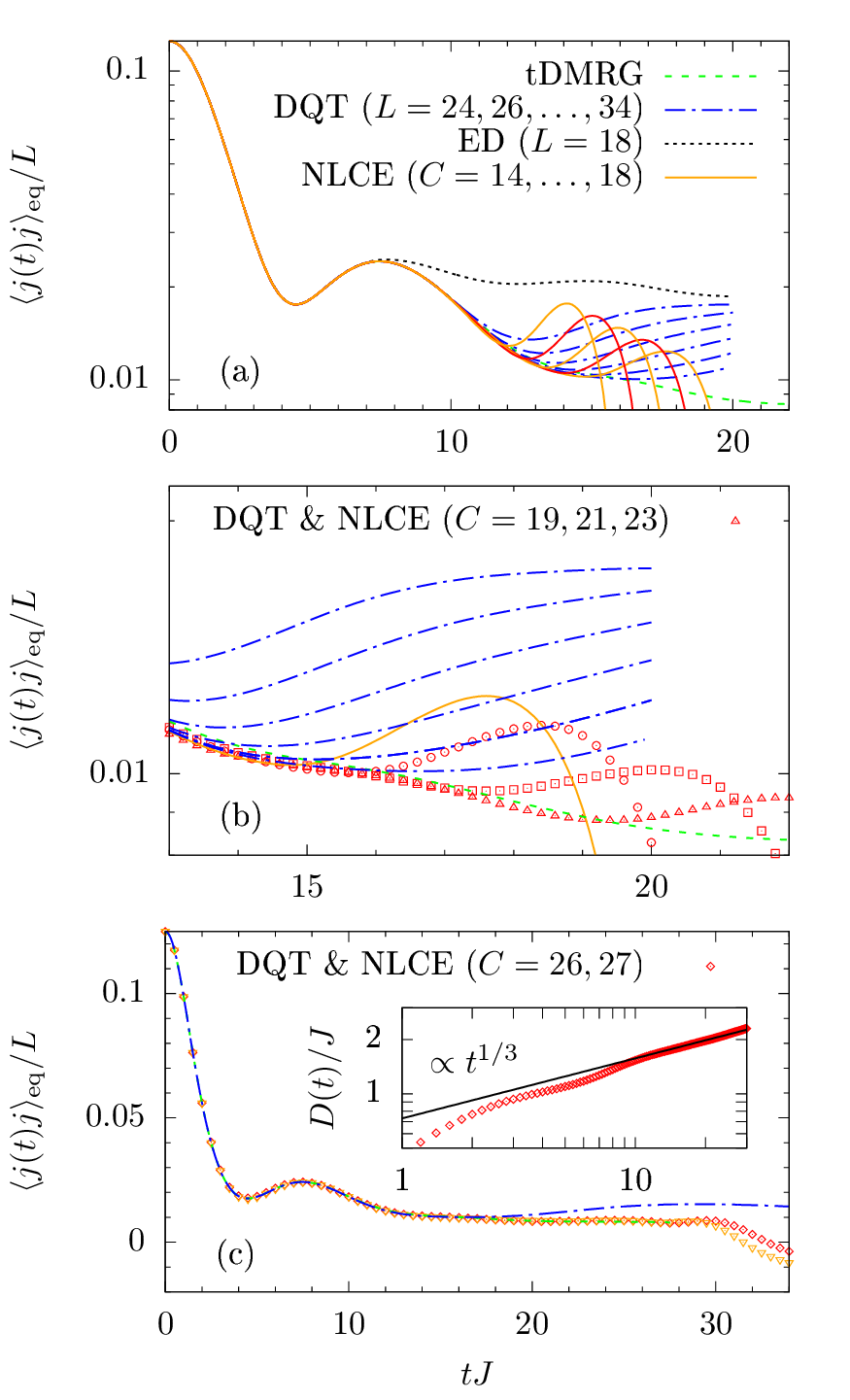}
\caption{(Color online) $\langle j(t) j 
\rangle_\text{eq}/L$ for $\Delta = 1$, $\Delta' = 0$, and  $\beta = 0$. (a) 
NLCE for $C = 14,\dots,18$, compared to ED, DQT \cite{steinigeweg2014,  
steinigeweg2015, Note2}, and tDMRG \cite{Karrasch2015}. (b) Combination DQT \& 
NLCE for $C = 19,21,23$. (c) Linear plot with data for $C = 26, 27$. Inset: 
Diffusion coefficient $D(t)$ for $C = 27$ and a power law $\propto t^{1/3}$ as a 
guide to the eye.}
\label{Fig2}
\end{figure}
In order to quantify $\epsilon$, Fig.\ \ref{Fig1}~(b) shows the absolute 
difference between ED and DQT for $L = 17$. For all times depicted, we find
$\epsilon = {\cal O}(10^{-5} - 10^{-6})$ when using $N_S = 5000$.

We now turn to our NLCE results, focusing on the integrable case $\Delta' = 0$ 
at infinite temperature $\beta = 0$. In Fig.\ \ref{Fig2}~(a), $\langle j(t) 
j\rangle_\text{eq}/L$ is shown at the isotropic point $\Delta = 1$, in a 
semilogarithmic plot. On the one hand, we depict NLCE data for $C = 
14, \dots, 18$. On the other hand, this data is compared to data for systems 
with periodic boundary conditions, obtained by either ED ($L = 18$) or DQT ($L 
= 24,26,\dots,34$) \cite{steinigeweg2014, steinigeweg2015}, and to tDMRG data 
\cite{Karrasch2015}. We find that NLCE converges up to a maximum time, 
increasing with $C$, until the expansion eventually breaks down. Remarkably, 
however, one can clearly see that NLCE is converged for a substantially longer 
time compared to ED, which becomes particularly evident when comparing NLCE for 
$C=18$ and ED for $L=18$. Moreover, NLCE for $C = 18$ coincides with tDMRG up 
to a time $t J \lesssim 15$, which approximately corresponds to the convergence 
reached in DQT with $L = 34$. This observation is particularly remarkable since 
the largest cluster in an expansion up to $C = 18$ also consists of $L = 18$ 
sites only. Thus, on short to intermediate times, NLCE clearly outperforms 
standard finite-size scaling.

\begin{figure}[tb]
\centering
\includegraphics[width=0.85\columnwidth]{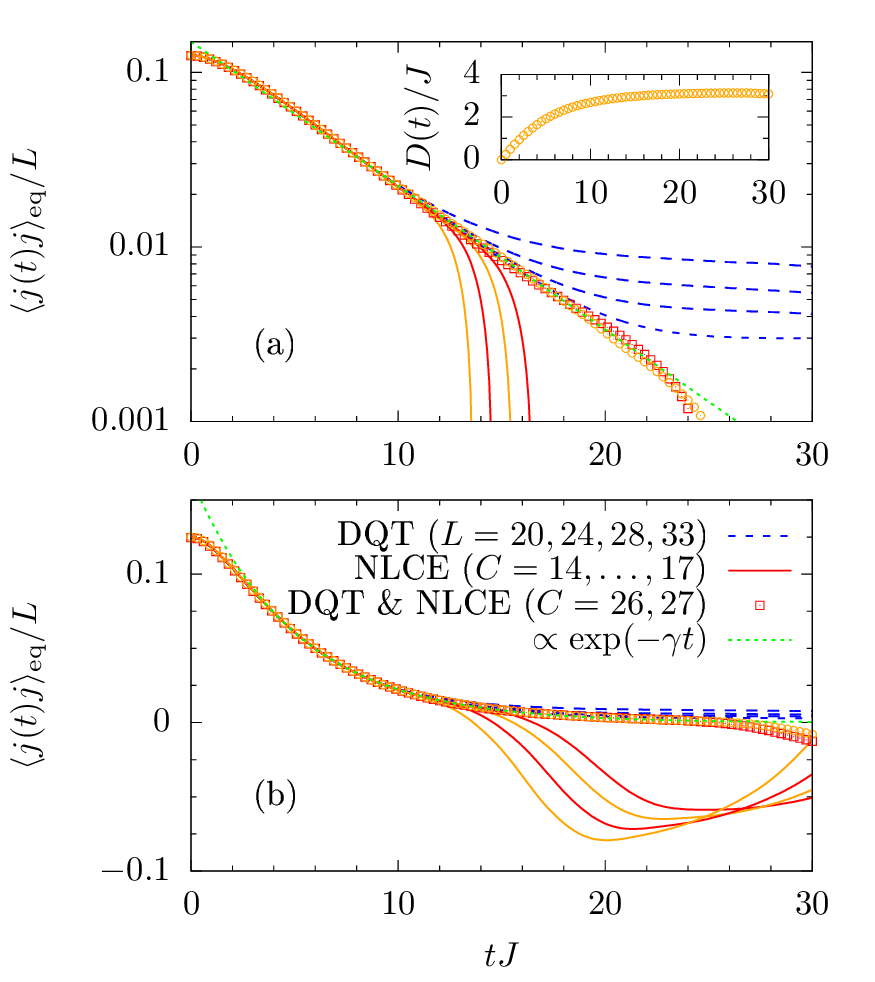}
\caption{(Color online) $\langle j(t) j \rangle_\text{eq}/L$ for $\Delta = 
\Delta' = 0.5$ and $\beta = 0$, obtained from NLCE ($C = 14,\dots, 17$) and DQT 
\& NLCE ($C = 26,27$) in a (a) semilogarithmic and (b) linear plot. An 
exponential is shown as a guide to the eye. Inset: Diffusion coefficient $D(t)$ 
for $C = 27$.}
\label{Fig3}
\end{figure}

Next, Fig.\ \ref{Fig2}~(b) shows a close-up of the data for intermediate times 
$13 \leq tJ \leq 22$. Furthermore, we now depict NLCE for larger $C = 19,21,23$, 
which can be obtained thanks to the combination DQT \& NLCE. For $tJ \leq 15$, 
one observes that all expansion orders lie on top of each other, demonstrating 
that DQT is indeed accurate enough to be combined with NLCE. Moreover, we find 
that NLCE continues to follow the tDMRG for longer and longer times when $C$ is 
increased. In fact, as can be seen in the linear plot in Fig.\ \ref{Fig2}~(c), 
NLCE for $C = 27$ essentially agrees with tDMRG up to times $tJ \lesssim 27$, in 
contrast to DQT data for $L = 34$ \cite{Note2}, which exhibits visible 
finite-size effects. Note that it is in principle possible to calculate even 
larger $c > 27$ \cite{super}.

Eventually, the inset in Fig.\ \ref{Fig2}~(c) shows the diffusion coefficient 
\cite{Steinigeweg2009}
\begin{equation}
D(t) = \frac{1}{\chi}\int_0^t \frac{\langle j(t') j \rangle}{L}\ \text{d}t'\ ,  
\quad \chi=\frac{1}{4}
\end{equation}
for the largest expansion order $C=27$. After $t J \sim 10$, $D(t)$ 
is consistent with a power-law scaling $\propto t^{1/3}$ and thus superdiffusive 
transport. This specific exponent has been recently suggested in Ref.\ 
\cite{Gopalakrishnan2018} (see also \cite{Ljubotina2017}), which cannot be 
seen in ED of small systems.

Finally, we are going to discuss NLCE for a nonintegrable model as well. In 
Fig.~\ref{Fig3}~(a), we show $\langle j(t)j\rangle_\text{eq}/L$ for $\Delta = 
\Delta' = 0.5$. Once again we compare NLCE for various $C$ to data for periodic 
chains of finite length \cite{Richter2018}. As a guide to the eye, we depict an 
exponential $\propto \exp(-\gamma t)$, which describes the decay process 
reasonably well \cite{Steinigeweg2011_2, Herbrych2012}. Similarly to 
Fig.~\ref{Fig2}, we find that NLCE outperforms standard finite-size scaling 
on short to medium time scales, in the sense that NLCE converges fast to the 
exponential even for small $C$. However, since finite-size effects are 
typically smaller for nonintegrable models, the advantage of NLCE becomes less 
pronounced compared to the integrable case shown in Fig.\ \ref{Fig2}. 


\section{Conclusion} \label{conclusion}

To summarize, we have shown that NLCE is a powerful approach 
to dynamics in a one-dimensional geometry, particularly when it is 
additionally combined with DQT. This we have done by comparing to existing 
results from various state-of-the-art methods. While we have focused on two case 
studies, the combination DQT \& NLCE yields equally convincing results for a 
wider choice of parameters and also at finite temperature $\beta > 0$ (see 
Appendix \ref{appendix_Delta} and \ref{appendix_FT} for details).

Promising directions of future research also include the application of 
the DQT \& NLCE combination to other dynamical quantities in one or two spatial
dimensions. In particular, the extension of NLCE to larger cluster sizes is 
of direct relevance to the study of quench dynamics starting from thermal 
initial states, cf.\ Ref.\ \cite{Mallayya2018}. Such classes of initial states 
have been shown to be amenable to the concept of typicality as well 
\cite{Richter2018_2, Richter2018_3}.


\subsection*{Acknowledgments}

We are grateful to C. Karrasch for sending us tDMRG data for comparison as well 
as to F.\ Heidrich-Meisner, L.\ Vidmar, J.\ Herbrych, and J.\ Gemmer for 
fruitful discussions. This work has been funded by the Deutsche 
Forschungsgemeinschaft (DFG) - STE 2243/3-1; 397067869; 355031190 - within the 
DFG Research Unit FOR 2692.

\begin{figure}[t]
\centering
\includegraphics[width=0.45\textwidth]{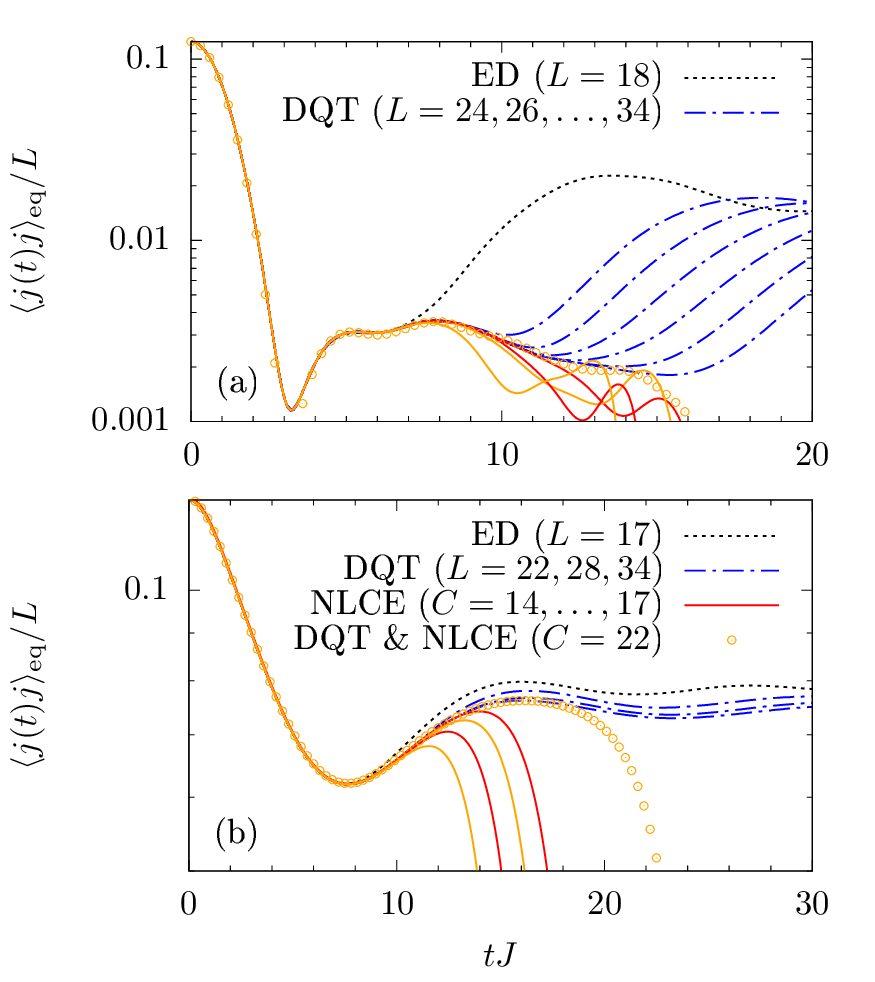}
\caption{(Color online) Current autocorrelation $\langle j(t) j 
\rangle_\text{eq}/L$  at $\beta = 0$ and anisotropy (a) $\Delta = 1.5$, 
$\Delta' = 0$; (b) $\Delta = 0.5$, $\Delta' = 0$. NLCE and combination DQT \& 
NLCE with different expansion order $C$, compared to ED and DQT calculations 
for finite systems with periodic boundary conditions \cite{steinigeweg2014,  
steinigeweg2015}.}
\label{FigS1}
\end{figure}
\begin{figure}[b]
\centering
\includegraphics[width=0.45\textwidth]{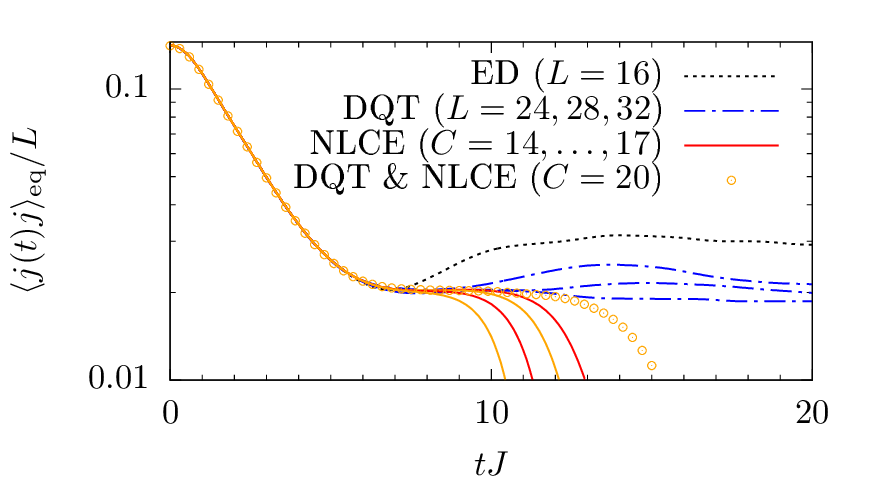}
\caption{(Color online) Current autocorrelation $\langle j(t) j 
\rangle_\text{eq}/L$ at finite temperature $\beta J = 1$ for $\Delta = 1$, 
$\Delta' = 0$. NLCE and combination DQT \& NLCE with different expansion order 
$C$, compared to ED and DQT calculations for finite systems with periodic 
boundary conditions \cite{steinigeweg2014, steinigeweg2015}.}
\label{FigS2}
\end{figure}

\appendix

\section{Current autocorrelations with NLCE for anisotropies $\Delta \neq 1$} 
\label{appendix_Delta}

In the main part of this paper, we have mostly focused on the isotropic 
Heisenberg chain with $\Delta = 1$. In order to substantiate our findings
even further, let us now present NLCE results also for other anisotropies  
$\Delta \neq 1$. In Fig.\ \ref{FigS1}, $\langle j(t)j\rangle_\text{eq}/L$ is 
shown for $\Delta = 1.5$ and $\Delta = 0.5$. In both cases, we again compare 
NLCE for expansion order $C = 14,\dots,17$ with ED and DQT for periodic chains 
of finite length. Comparing small and large $\Delta$, we find that the 
convergence of the NLCE is a little slower in the case of $\Delta = 1.5$, where 
the expansion also exhibits even-odd-like effects. This might be caused by the 
fact that the autocorrelation function takes on very small values for this 
choice of $\Delta$. Generally, however, the situation for $\Delta = 1.5$ and 
$\Delta = 0.5$ is very similar compared to the isotropic case discussed 
in the main part of this paper, i.e., NLCE for given expansion order is 
converged to the thermodynamic limit for a longer time scale than in 
corresponding calculations of systems with periodic boundaries.

\begin{table}[t]
\caption{Number of random states $N_S$ for a given system size $L$, as used in  
the DQT calculations underlying the NLCE for $\Delta = 1$, $\Delta'=0$, and 
$\beta = 0$ in Figs.\ \ref{Fig2}~(b) and (c). The product $N_S \times 2^L$ is 
always larger than $5000 \times 2^{17}$.}
\begin{tabular}{|r|r|r|}
\hline
$L$ & $N_S$ & $N_S \times 2^L$ \\
\hline \hline
$ 17 $ & $5000$ & $6.55 \times 10^8$ \\
\hline
\hline
$ 19 $ & $7900$ & $41.4 \times 10^8$ \\
$ 20 $ & $3725$ & $39.1 \times 10^8$ \\
$ 21 $ & $1260$ & $26.4 \times 10^8$ \\
$ 22 $ &  $643$ & $27.0 \times 10^8$ \\
$ 23 $ &  $298$ & $25.0 \times 10^8$ \\
$ 24$  &   $85$ & $14.3 \times 10^8$ \\
$ 25$  &   $44$ & $14.8 \times 10^8$ \\
$ 26$  &   $34$ & $22.8 \times 10^8$ \\
$ 27$  &    $8$ & $10.7 \times 10^8$ \\
\hline 
\end{tabular}
\label{tab1}
\end{table}
\begin{figure}[t]
\centering
\includegraphics[width=0.45\textwidth]{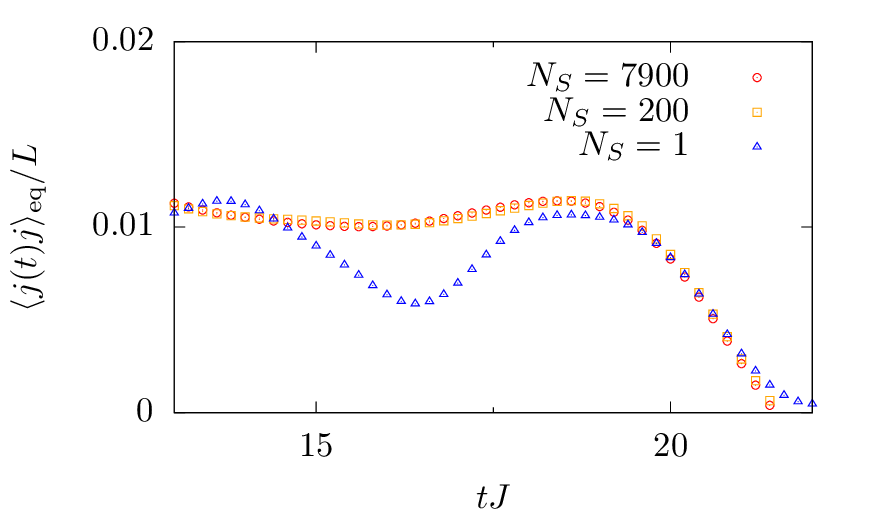}
\caption{(Color online) Expansion order $C=19$ according to the combination 
DQT \& NLCE, as shown in Fig.\ \ref{Fig2}~(b) in the main text, but now for a 
different number of random states: $N_S = 1$, $200$, and $7900$. While the
curves for $N_S = 200$ and $7900$ are practically the same, $N_S = 1$ clearly 
differs for such a small $C$.}
\label{Fig6}
\end{figure}

\section{Current autocorrelations with NLCE for finite temperatures} 
\label{appendix_FT}

Eventually, let us demonstrate that NLCE is certainly not restricted to the 
infinite-temperature limit. To this end, Fig.\ \ref{FigS2} exemplarily shows 
$\langle j(t)j\rangle_\text{eq}/L$ at the finite temperature $\beta J = 1$ for 
the single choice $\Delta = 1$. We find that, already for $C = 20$, NLCE is 
able to see a constant plateau up to times $tJ \lesssim 12$, which is 
clearly missed by ED for $L = 16$ and only captured by DQT for significantly 
larger systems with $L = 28, 32$. Thus, we conclude that NLCE also provides a 
powerful approach to current autocorrelations for a wider range of temperatures.

\section{Number of random states} \label{appendix_rs}

In Tab.\ \ref{tab1}, we specify the number of random states $N_S$ for a given 
system size $L$, as used in the DQT calculations underlying the NLCE for 
$\Delta = 1$, $\Delta'=0$, and $\beta = 0$ in Figs.\ \ref{Fig2}~(b) and (c). As 
stated before, the product $N_S \times 2^L$ is always larger than 
$5000 \times 2^{17}$. While it is evident from the comprehensive comparison in 
the main text that this averaging is sufficient, it might also be insightful to 
see that averaging is indeed important for small expansion orders $C$. We 
therefore depict in Fig.\ \ref{Fig6} the expansion order $C=19$ according to 
the combination DQT \& NLCE, as shown in Fig.\ \ref{Fig2}~(b), but now for a 
different number of random states: $N_S = 1$, $200$, and $7900$. While the
curves for $N_S = 200$ and $7900$ are practically the same, $N_S = 1$ clearly 
differs for such a small $C$.


\end{document}